\documentclass[runningheads,a4paper]{llncs}

\usepackage[english]{babel}
\usepackage{amsfonts}
\usepackage{amsmath}
\usepackage{amssymb}
\usepackage{latexsym}
\usepackage{mathrsfs}
\usepackage{psfrag}
\usepackage{color}
\usepackage[applemac]{inputenc}

\usepackage{amssymb}
\usepackage{graphicx}

\usepackage{url}
\urldef{\mailsa}\path|{marco.liuni, axel.roebel, xavier.rodet}@ircam.fr|
\urldef{\mailsb}\path|marco.romito@math.unifi.it|
\newcommand{\keywords}[1]{\par\addvspace\baselineskip
\noindent\keywordname\enspace\ignorespaces#1}

\newcommand{\m}{\mathbb}

\newcommand{\ca}{\hat}
\newcommand{\de}{\mathrm{d}}

\newcommand{\beq}{\begin{equation}}
\newcommand{\eeq}{\end{equation}}
\newcommand{\lsc}{\langle}
\newcommand{\rsc}{\rangle}

\begin{document}

\mainmatter  % start of an individual contribution

\title{An Entropy Based Method for Local Time-adaptation of the Spectrogram}

\titlerunning{A Method for Local Time-adaptation of the Spectrogram}

\author{Marco Liuni$_{1,2}$ \thanks{This work is supported by grants from Region Ile-de-France.}%%
\and Axel R\"obel$_2$ \and Marco Romito$_1$ \and Xavier Rodet$_2$
}

\authorrunning{A Method for Local Time-adaptation of the Spectrogram}

\institute{$^1$Università di Firenze, Dip. di Matematica "U. Dini"\\ Viale Morgagni, 67/a - 50134 Florence - ITALY\\$^{2}$ IRCAM - CNRS STMS, Analysis/Synthesis Team\\1, place Igor-Stravinsky - 75004 Paris - FRANCE\\ \vspace{0.5cm}
\mailsa\\
\mailsb\\
\url{http://www.ircam.fr/anasyn.html}}


\toctitle{Lecture Notes in Computer Science}
\tocauthor{Authors' Instructions}
\maketitle

\begin{abstract}
We propose a method for automatic local time-adaptation of the spectrogram of audio signals: it is based on the decomposition of a signal within a Gabor multi-frame through the STFT operator. The sparsity of the analysis in every individual frame of the multi-frame is evaluated through the Rényi entropy measures: the best local resolution is determined minimizing the entropy values. The overall spectrogram of the signal we obtain thus provides local optimal resolution adaptively evolving over time.
We give examples of the performance of our algorithm with an instrumental sound and a synthetic one, showing the improvement in spectrogram displaying obtained with an automatic adaptation of the resolution. The analysis operator is invertible, thus leading to a perfect reconstruction of the original signal through the analysis coefficients.
\keywords{adaptive spectrogram, sound representation, sound analysis, sound synthesis, Rényi entropy, sparsity measures, frame theory}
\end{abstract}

\section{Introduction}

Far from being restricted to entertainment, sound processing techniques are required in many different domains: they find applications in medical sciences, security instruments, communications among others. The most challenging class of signals to consider is indeed music: the completely new perspective opened by contemporary music, assigning a fundamental role to concepts as noise and timbre, gives musical potential to every sound.\\

The standard techniques of digital analysis are based on the decomposition of the signal in a system of elementary functions, and the choice of a specific system necessarily has an influence on the result. Traditional methods based on single sets of atomic functions have important limits: a Gabor frame imposes a fixed resolution over all the time-frequency plane, while a wavelet frame gives a strictly determined variation of the resolution: moreover, the user is frequently asked to define himself the analysis window features, which in general is not a simple task even for experienced users. This motivates the search for adaptive methods of sound analysis and synthesis, and for algorithms whose parameters are designed to change according to the analyzed signal features. Our research is focused on the development of mathematical models and tools based on the local automatic adaptation of the system of functions used for the decomposition of the signal: we are interested in a complete framework for analysis, spectral transformation and re-synthesis; thus we need to define an efficient strategy to reconstruct the signal through the adapted decomposition, which must give a perfect recovery of the input if no transformation is applied.\\

Here we propose a method for local automatic time-adaptation of the Short Time Fourier Transform window function, through a minimization of the \emph{Rényi entropy} \cite{ZZ97} of the spectrogram; we then define a re-synthesis technique with an extension of the method proposed in \cite{GL84}. Our approach can be presented schematically in three parts:

\begin{enumerate}

\item a model for signal analysis exploiting concepts of Harmonic Analysis, and Frame Theory in particular: it is a generally highly redundant decomposing system belonging to the class of multiple Gabor frames \cite{Do02},\cite{JBD09};\\

\item a sparsity measure defined on time-frequency localized subsets of the analysis coefficients, in order to determine local optimal concentration;\\

\item a reduced representation obtained from the original analysis using the information about optimal concentration, and a synthesis method through an expansion in the reduced system obtained.

\end{enumerate}

We have realized a first implementation of this scheme in two different versions: for both of them a sparsity measure is applied on subsets of analysis coefficients covering the whole frequency dimension, thus defining a time-adapted analysis of the signal. The main difference between the two concerns the first part of the model, that is the single frames composing the multiple Gabor frame. This is a key point as the first and third part of the scheme are strictly linked: the frame used for re-synthesis is a reduction of the original multi-frame, so the entire model depends on how the analysis multi-frame is designed. The section \emph{Frame Theory in Sound Analysis and Synthesis} treats this part of our research in more details.\\

The second point of the scheme is related to the measure applied on the coefficients of the analysis within the multi-frame to determine local best resolutions. We consider measures borrowed from Information Theory and Probability Theory according to the interpretation of the analysis within a frame as a probability density \cite{Co89}: our model is based on a class of entropy measures known as \emph{Rényi entropies} which extend the classical Shannon entropy. The fundamental idea is that minimizing the complexity or information over a set of time-frequency representations of the same signal is equivalent to maximizing the concentration and peakiness of the analysis, thus selecting the best resolution tradeoff \cite{BF01}: in the section \emph{Rényi Entropy of Spectrograms} we describe how a sparsity measure can consequently be defined through an information measure. Finally, in the fourth section we provide a description of our algorithm and examples of adapted spectrogram for different sounds.\\

Some examples of this approach can be found in the literature: the idea of gathering a sparsity measure from Rényi entropies is detailed in \cite{BF01}, and in \cite{Ja05} a local time-frequency adaptive framework is presented exploiting this concept, even if no methods for perfect reconstruction are provided. In \cite{WGD01} sparsity is obtained through a regression model; a recent development in this sense is contained in \cite{JBD09} where a class of methods for analysis adaptation are obtained separately in the time and frequency dimension together with perfect reconstruction formulas: indeed no strategies for automatization are employed, and adaptation has to be managed by the user. The model conceived in \cite{RPW09} belongs to this same class but presents several novelties in the construction of the Gabor multi-frame and in the method for automatic local time-adaptation. In \cite{LT06} another time-frequency adaptive spectrogram is defined considering a sparsity measure called \emph{energy smearing}, without taking into account the re-synthesis task. The concept of \emph{quilted frame}, recently introduced in \cite{Do10}, is the first promising effort to establish a unified mathematical model for all the various frameworks cited above.

\section{Frame Theory in Sound Analysis and Synthesis}

When analyzing a signal through its decomposition, the features of the representation are influenced by the decomposing functions; the Frame Theory (see \cite{Ch03},\cite{Gr01} for detailed mathematical descriptions) allows a unified approach when dealing with different bases and systems, studying the properties of the operators that they identify. The concept of frame extends the one of orthonormal basis in a Hilbert space, and it provides a theory for the discretization of time-frequency densities and operators \cite{Do02}, \cite{Sun09}, \cite{BGL10}. Both the STFT and the Wavelet transform can be interpreted within this setting (see \cite{Mal99} for a comprehensive survey of theory and applications).\\
Here we summarize the basic definitions and theorems, and outline the fundamental step consisting in the introduction of \emph{Multiple Gabor Frames}, which is comprehensively treated in \cite{Do02}. The problem of standard frames is that the decomposing atoms are defined from the same original function, thus imposing a limit on the type of information that one can deduce from the analysis coefficients; if we were able to consider frames where several families of atoms coexist, than we would have an analysis with variable information, at the price of a higher redundancy.

\subsection{Basic Definitions and Results}
Given a Hilbert space $H$ seen as a vector space on $\m{C}$, with its own scalar product, we consider in $H$ a set of vectors $\{\phi_{\gamma}\}_{\gamma \in \Gamma}$ where the index set $\Gamma$ may be infinite and $\gamma$ can also be a multi-index. The set $\{\phi_{\gamma}\}_{\gamma\in \Gamma}$ is a \emph{frame} for $H$ if there exist two positive non zero constants $A$ and $B$, called \emph{frame bounds}, such that for all $f\in H$, 
\begin{equation}\label{frame_complete}A\|f\|^2\leq\sum_{\gamma\in\Gamma}|\lsc f,\phi_{\gamma}\rsc|^2\leq B\|f\|^2~.\end{equation}
We are interested in the case $H = L^2(\m{R})$ and $\Gamma$ countable, as it represents the standard situation where a signal $f$ is decomposed through a countable set of given functions $\{\phi_{k}\}_{k\in \m{Z}}$. The frame bounds $A$ and $B$ are the infimum and supremum, respectively, of the eigenvalues of the \emph{frame operator} $\mathrm{U}$, defined as
\beq\label{frame_op_def} \mathrm{U}f = \sum_{k\in\m{Z}} \lsc f,\phi_{k}\rsc \phi_{k}~.
\eeq
For any frame $\{\phi_{k}\}_{k\in \m{Z}}$ there exist dual frames $\{\tilde{\phi}_{k}\}_{k\in \m{Z}}$ such that for all $f\in L^2(\m{R})$
\beq\label{dual_frame_def} f = \sum_{k\in\m{Z}} \lsc f,\phi_{k}\rsc \tilde{\phi}_{k} = \sum_{k\in\m{Z}} \lsc f,\tilde{\phi}_{k}\rsc \phi_{k} ~,
\eeq
so that given a frame it is always possible to perfectly reconstruct a signal $f$ using the coefficients of its decomposition through the frame. The inverse of the frame operator allows the calculation of the canonical dual frame 
\beq\label{dual_frame_def} \tilde{\phi}_{k} =\mathrm{U}^{-1} \phi_{k}~\eeq 
which guarantees minimal-norm coefficients in the expansion.\\

A \emph{Gabor frame} is obtained by time-shifting and frequency-transposing a window function $g$ according to a regular grid. They are particularly interesting in the applications as the analysis coefficients are simply given by sampling the STFT of $f$ with window $g$ according to the nodes of a specified lattice. Given a time step $a$ and a frequency step $b$ we write $\{u_n\}_{n\in\m{Z}} = an$ and $\{\xi_k\}_{k\in\m{Z}} = bk$; these two sequences generate the nodes of the time-frequency lattice $\Lambda$ for the frame $\{g_{n,k}\}_{(n,k)\in\m{Z}^2}$ defined as
\beq\label{gabor_frame_def} g_{n,k}(t) = g(t - u_n)e^{2\pi i\xi_k t}~;
\eeq
the nodes are the centers of the Heisenberg boxes associated to the windows in the frame. The lattice has to satisfy certain conditions for $\{g_{n,k}\}$ to be a frame \cite{Da90}, which impose limits on the choice of the time and frequency steps: for certain choices \cite{Da86} which are often adopted in standard applications, the frame operator takes the form of a multiplication,
\beq\label{painless_op_def} \mathrm{U}f(t) = \left(b^{-1}\sum_{n\in\m{Z}} |g(t -u_n)|^2\right) f(t)~,
\eeq
and the dual frame is easily calculated by means of a straight multiplication of the atoms in the original frame. The relation between the steps $a$, $b$ and the frame bounds $A$, $B$ in this case is clear by \eqref{painless_op_def}, as the frame condition implies
\beq\label{painless_bounds} 0 < A \leq b^{-1}\sum_{n\in\m{Z}} |g(t -u_n)|^2 \leq B < \infty~.
\eeq
Thus we see that the frame bounds provide also information on the redundancy of the decomposition of the signal within the frame.\\ 

\subsection{Multiple Gabor Frames}
In our adaptive framework, we look for a method to achieve an analysis with multiple resolutions: thus we need to combine the information coming from the decompositions of a signal in several frames of different window functions. Multiple Gabor frames have been introduced in \cite{ZZ97} to provide the original Gabor analysis with flexible multi-resolution techniques: given a set of index $L\subseteq \m{Z}$ and different frames $\{g^l_{n,k}\}_{(n,k)\in\m{Z}^2}$ with $l\in L$, a multiple Gabor frame is obtained with a union of the single given frames. The different $g^l$ do not necessarily share the same type or shape: in our method an original window is modified with a finite number of scaling
\beq\label{scaling} g^l(t) = \frac{1}{\sqrt{l}}~g\bigg(\frac{t}{l}\bigg)~; \eeq
then all the scaled versions are used to build $|L|$ different frames which constitute the initial multi-frame.\\ 

A Gabor multi-frame has in general a significant redundancy which lowers the readability of the analysis. A possible strategy to overcome this limit is proposed in \cite{JBD09} where \emph{nonstationary Gabor frames} are introduced, actually allowing the choice of a different window for each time location of a  global irregular lattice $\Lambda$, or alternatively for each frequency location. This way, the window chosen is a function of time or frequency position in the time-frequency space, not both. In most applications, for this kind of frame there exist fast FFT based methods for the analysis and re-synthesis steps. Referring to the time case, with the abuse of notation $g_{n(l)}$ we indicate the window $g^l$ centered at a certain time $n(l) = u_n$ which is a function of the chosen window itself. Thus, a nonstationary Gabor frame is given by the set of atoms
\beq\label{nonstat_frame_def} \{g_{n(l)} e^{2\pi i b_l k t}, (n(l),b_l k) \in \Lambda \}~,\eeq 
where $b_l$ is the frequency step associated to the window $g^l$ and $k\in \m{Z}~$. If we suppose that the windows $g^l$ have limited time support and a sufficiently small frequency step $b_l$, the frame operator $\mathrm{U}$ takes a similar form to the one in \eqref{painless_op_def},
\beq\label{frame_op_diag} \mathrm{U}f(t) = \sum_{n(l)} \frac{1}{b_l} |g_{n(l)}(t)|^2f(t)~.\eeq
Here, if $ N = \sum_l \frac{1}{b_l} |g_{n(l)}(s)|^2 \simeq 1$ then $\mathrm{U}$ is invertible and the set \eqref{nonstat_frame_def} is a frame whose dual frame is given by
\beq\label{dual_frame_nonst} \tilde{g}_{n(l),k}(t) = \frac{1}{N} g_{n(l)}(t) e^{2\pi i b_l k t}~.\eeq

Nonstationary Gabor frames belong to the recently introduced class of \emph{quilted frames} \cite{Do10}: in this kind of decomposing systems the choice of the analysis window depends on both the time and the frequency location, causing more difficulties for an analytic fast computation of a dual frame as in \eqref{dual_frame_nonst}: future improvements of our research concern the employment of such a decomposition model for automatic local adaptation of the spectrogram resolution both in the time and the frequency dimension.

\section{Rényi Entropy of Spectrograms}

We consider the discrete spectrogram of a signal as a sampling of the square of its continuous version
\beq\label{spec_def}  \mathrm{PS}_f(u,\xi) = |\mathrm{S}f(u,\xi)|^2 = \bigg |\int f(t)g(t-u)e^{-2 \pi i\xi t}\de t\bigg |^2~,\eeq
where $f$ is a signal, $g$ is a window function and $\mathrm{S}f(u,\xi)$ is the STFT of $f$ through $g$.\\
Such a sampling is obtained according to a regular lattice $\Lambda_{ab}$, considering a Gabor frame \eqref{gabor_frame_def},
\beq\label{disc_spec_def}  \mathrm{PS}_f[n,k] = |\mathrm{S}f[u_n,\xi_k]|^2~.\eeq
 
With an appropriate normalization both the continuous and discrete spectrogram can be interpreted as probability densities. Thanks to this interpretation, some techniques belonging to the domain of Probability and Information Theory can be applied to our problem: in particular, the concept of \emph{entropy} can be extended to give a sparsity measure of a time-frequency density. A promising approach \cite{BF01} takes into account Rényi entropies, a generalization of the Shannon entropy: the application to our problem is related to the concept that minimizing the complexity or information of a set of time-frequency representations of a same signal is equivalent to maximizing 
the concentration, peakiness, and therefore the sparsity of the analysis. Thus we will consider as \emph{best} analysis the sparsest one, according to the minimal entropy evaluation.\\

Given a signal $f$ and its spectrogram $\mathrm{PS}_f$ as in \eqref{spec_def}, the \emph{Rényi entropy} of \emph{order} $\alpha > 0,~\alpha \neq 1$ of $\mathrm{PS}_f$ is defined as follows
\beq\label{ren_ent_def}  \mathrm{H}_{\alpha}^R(\mathrm{PS}_f) =\frac{1}{1-\alpha}~\log_2 \iint_R \bigg(\frac{\mathrm{PS}_f (u,\xi)}{\iint_R\mathrm{PS}_f(u',\xi')\de u'\de\xi'}\bigg)^{\alpha}\de u \de \xi~, \eeq
where $R\subseteq \m{R}^2$ and we omit its indication if equality holds. Given a discrete spectrogram with time step $a$ and frequency step $b$ as in \eqref{disc_spec_def}, we consider $R$ as a rectangle of the time-frequency plane $R = [t_1,t_2]\times[\nu_1,\nu_2] \subseteq \m{R}^2$. It identifies a sequence of points $G \subseteq \Lambda_{ab}$ where $G = \{(n,k)\in\m{Z}^2: t_1 \leq na \leq t_2,~ \nu_1 \leq kb \leq \nu_2\}$. As a discretization of the original continuous spectrogram, every sample in $\mathrm{PS}_f[n,k]$ is related to a time-frequency region of area $ab$; we thus obtain the discrete Rényi entropy measure directly from \eqref{ren_ent_def},
\beq\label{ren_ent_disc} \mathrm{H}_{\alpha}^G[\mathrm{PS}_f ] =  \frac{1}{1-\alpha}\log_2 \sum_{[n,k]\in G} \bigg(\frac{\mathrm{PS}_f[n,k]}{\sum_{[n',k']\in G} \mathrm{PS}_f[n',k']}\bigg)^{\alpha}  + \log_2(ab)~.\eeq

We will focus on discretized spectrograms with a finite number of coefficients, as dealing with digital signal processing requires to work with finite sampled signals and distributions.\\

Among the general properties of Rényi entropies \cite{Re61}, \cite{BS93} and \cite{Zy04} we recall in particular those directly related with our problem. It is easy to show that for every finite discrete probability density $P$ the entropy $\mathrm{H}_{\alpha}(P)$ tends to coincide with the Shannon entropy of $P$ as the order $\alpha$ tends to one. Moreover, $\mathrm{H}_{\alpha}(P)$ is a non increasing function of $\alpha$, so
\beq\label{ren_ent_nonincalpha} \alpha_1 < \alpha_2 \Rightarrow \mathrm{H}_{\alpha_1}(P)\geq \mathrm{H}_{\alpha_2}(P)~.\eeq
As we are working with finite discrete densities we can also consider the case $\alpha = 0$ which is simply the logarithm of the number of elements in $P$; as a consequence $\mathrm{H}_0(P) \geq \mathrm{H}_{\alpha}(P)$ for every admissible order $\alpha$.\\
A third basic fact is that for every order $\alpha$ the Rényi entropy $\mathrm{H}_{\alpha}$ is maximum when $P$ is uniformly distributed, while it is minimum and equal to zero when $P$ has a single non-zero value.\\

All of these results give useful information on the values of different measures on a single density $P$ as in \eqref{ren_ent_disc}, while the relations between the entropies of two different densities $P$ and $Q$ are in general hard to determine analytically; in our problem, $P$ and $Q$ are two spectrograms of a signal in the same time-frequency area, based on two window functions with different scaling as in \eqref{scaling}. In some basic cases such a relation is achievable, as shown in the following example.

\subsection{Best Window for Sinusoids}
When the spectrograms of a signal through different window functions do not depend on time, it is easy to compare their entropies: let $\mathrm{PS}_s$ be the sampled spectrogram of a sinusoid $s$ over a finite region $G$ with a window function $g$ of compact support; then $\mathrm{PS}_s$ is simply a translation in the frequency domain of $\ca{g}$, the Fourier transform of the window, and it is therefore time-independent. We choose a bounded set $L$ of admissible scaling factors, so that the discretized support of the scaled windows $g^l$ still remains inside $G$ for any $l\in L$. It is not hard to prove that the entropy of a spectrogram taken with such a scaled version of $g$ is given by
 \beq\label{ren_ent_scaledspec} \mathrm{H}_{\alpha}^G(\mathrm{PS}_{sl}) = \mathrm{H}_{\alpha}^G(\mathrm{PS}_s ) - \log_2 l~.\eeq 
The sparsity measure we are using chooses as best window the one which minimizes the entropy measure: we deduce from \eqref{ren_ent_scaledspec} that it is the one obtained with the largest scaling factor available, therefore with the largest time-support. This is coherent with our expectation as stationary signals, such as sinusoids, are best analyzed with a high frequency resolution, because time-independency allows a small time resolution. Moreover, this is true for any order $\alpha$ used for the entropy calculus. Symmetric considerations apply whenever the spectrogram of a signal does not depend on frequency, as for impulses.\\

\subsection{The $\mathbb{\alpha}$ Parameter}
\label{ssec:sub_alpha}
\begin{figure}

\begin{minipage}[b]{1.0\linewidth}
  \centering
  \centerline{\includegraphics[width=11.5cm]{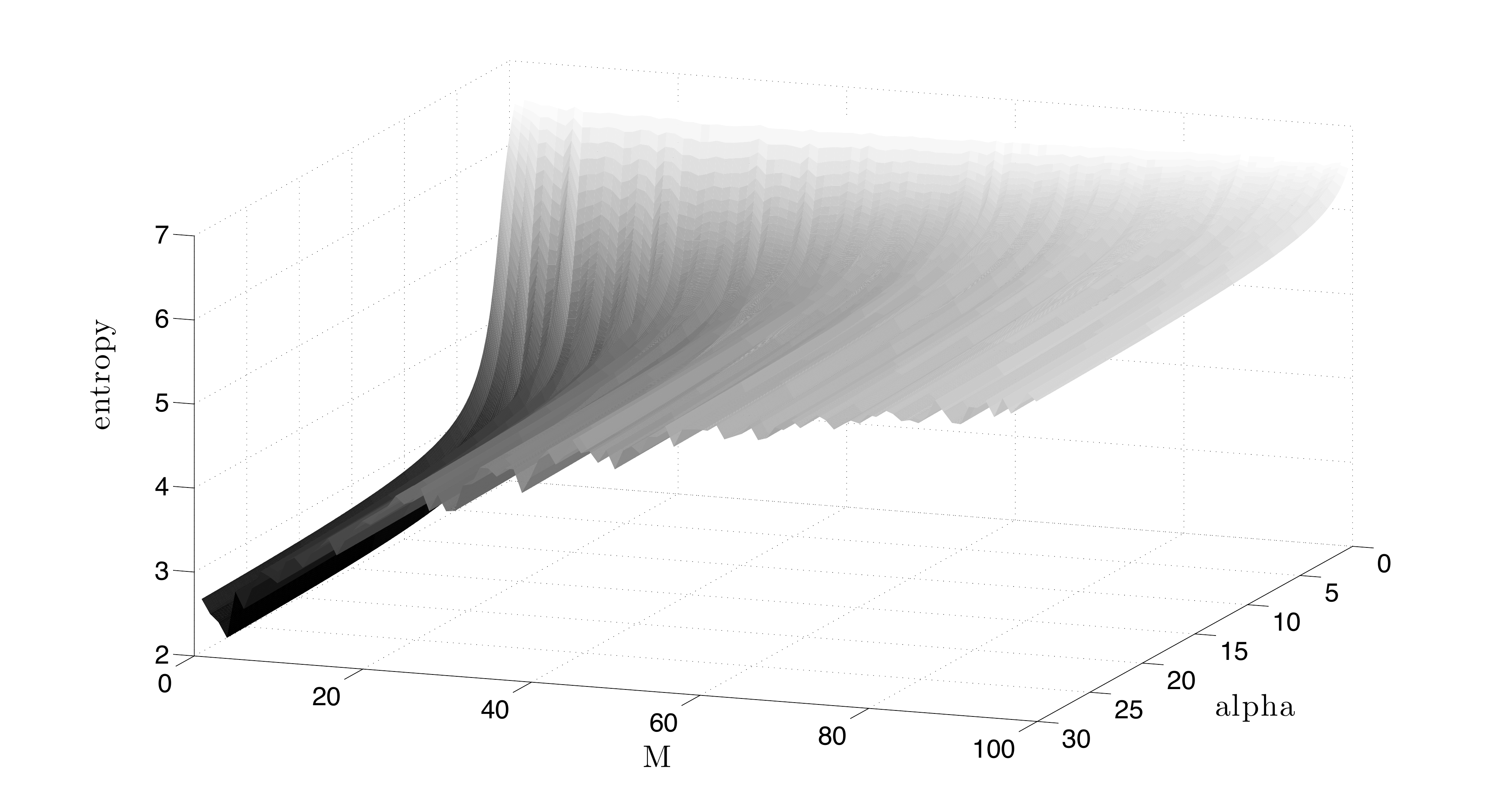}}\medskip
\vspace{-0.5cm}
\end{minipage}

\caption{Rényi entropy evaluations of the $D_M$ vectors with varying $\alpha$; the distribution becomes flatter as $M$ increases.}
\label{fig:test_alpha}
\end{figure}

The $\alpha$ parameter in \eqref{ren_ent_def} introduces a biasing on the spectral coefficients; to have a qualitative description of this biasing, we consider a collection of simple spectrograms composed by a variable amount of large and small coefficients. We realize a vector $D$ of length $N = 100$ generating numbers between 0 and 1 with a normal random distribution; then we consider the vectors $D_M,~1\leq M\leq N$ such that

\begin{displaymath} D_M[k] = \left\{ \begin{array}{ll}
D[k] & \textrm{if $k\leq M$}\\ \frac{D[k]}{20} & \textrm{if $k > M$}
\end{array} \right. \vspace{-8mm} \end{displaymath}
\beq\label{spec_models}\vspace{5mm}\eeq
and then normalize to obtain a unitary sum. We then apply Rényi entropy measures with $\alpha$ varying between 0 and 30: as we see from figure \ref{fig:test_alpha}, there is a relation between $M$ and the slope of the entropy curves for the different values of $\alpha$.
For $\alpha = 0$, $\mathrm{H}_0[D_M]$ is the logarithm of the number of non-zero coefficients and it is therefore constant; when $\alpha$ increases, we see that densities with a small amount of large coefficients gradually decrease their entropy, faster than the almost flat vectors corresponding to larger values of $M$. This means that by increasing $\alpha$ we emphasize the difference between the entropy values of a peaky distribution and that of a nearly flat one. The sparsity measure we consider select as best analysis the one with minimal entropy, so reducing $\alpha$ rises the probability of less peaky distributions to be chosen as sparsest: in principle, this is desirable as weaker components of the signal, such as partials, have to be taken into account in the sparsity evaluation. But as well, this principle should be applied with care as a small coefficient in a spectrogram could be determined by a partial as well as by a noise component; choosing an extremely small $\alpha$, the best window chosen could vary without a reliable relation with spectral concentration depending on the noise level within the sound.

\subsection{Time and Frequency Steps}
A last remark regards the dependency of \eqref{ren_ent_disc} on the time and frequency step $a$ and $b$ used for the discretization of the spectrogram. When considering signals as finite vectors, a signal and its Fourier Transform have the same length. Therefore in the STFT the window length determines the number frequency points, while the sampling rate sets frequency values: the definition of $b$ is thus implicit in the window choice. Actually, the FFT algorithm allows to specify a number of frequency points larger than the signal length: further frequency values are obtained as an interpolation between the original ones by properly adding zero values to the signal. If the sampling rate is fixed, this procedure causes a smaller $b$ as a consequence of a larger number of frequency points. We have numerically verified that such a variation of $b$ has no impact on the entropy calculus, so that the FFT size can be set according to implementation needs.\\
Regarding the time step $a$, we are working on the analytical demonstration of a largely verified evidence: as long as the decomposing system is a frame the entropy measure is invariant to redundancy variation, so the choice of $a$ can be ruled by considerations on the invertibility of the decomposing frame without losing coherence between the information measure of the different analyses. This is a key point, as it states that the sparsity measure obtained allows a total independence between the hop sizes of the different analyses: with the implementation of proper structures to handle multi-hop STFTs we have obtained a more efficient algorithm in comparison with those imposing a fixed hop size, as \cite{LT06} and the first version of the one we have realized.

\section{Algorithm and Examples}

We now summarize the main operations of  the algorithm we have developed providing examples of its application. For the calculation of the spectrograms we use a \emph{Hanning window}
\beq\label{han_def} h(t) = \cos^2(\pi t)\chi_{[-\frac{1}{2},\frac{1}{2}]}~,\eeq
with $\chi$ the indicator function of the specified interval, but it is obviously possible to generalize the results thus obtained to the entire class of compactly supported window functions. In both the versions of our algorithm we create a multiple Gabor frame as in \eqref{gabor_frame_def}, using as mother functions some scaled version of $h$, obtained as in \eqref{scaling} with a finite set of positive real scaling factors $L$.\\
We consider consecutive segments of the signal, and for each segment we calculate $|L|$ spectrograms with the $|L|$ scaled windows: the length of the analysis segment and the overlap between two consecutive segments are given as parameters.\\

\begin{figure}
\centerline{\includegraphics[width=11.5cm]{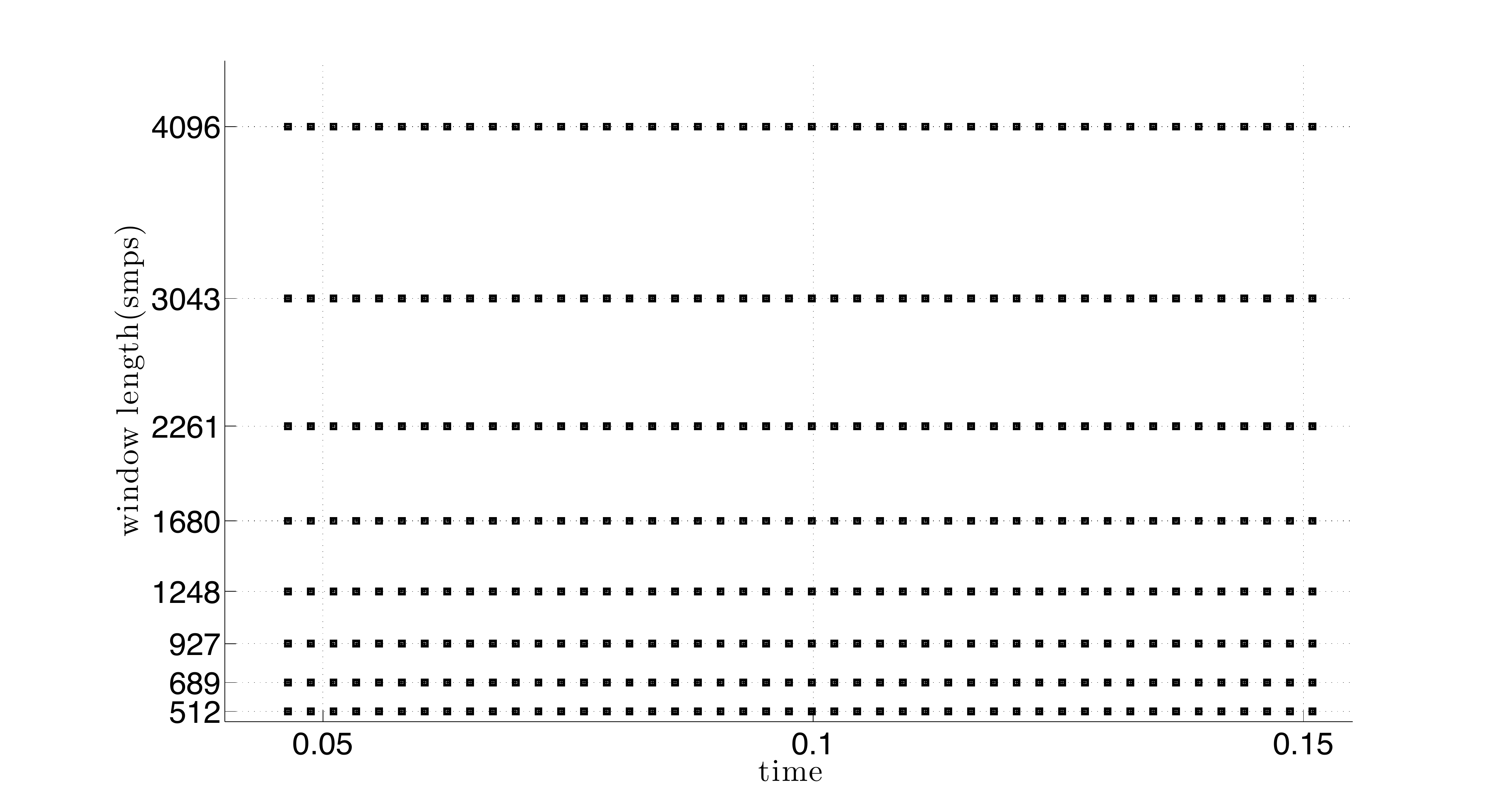}}
\caption[Figure 1]{\label{fig:centers_fix} \it An analysis segment: time locations of the Heisenberg boxes associated to the multi-frame used in the first version of our algorithm.}
\end{figure}

In the first version of the algorithm the different frames composing the multi-frame have the same time step $a$ and frequency step $b$: this guarantees that for each signal segment the different frames have Heisenberg boxes whose centers lay on a same lattice on the time-frequency plane, as illustrated in figure \ref{fig:centers_fix}. To guarantee that all the $|L|$ scaled windows constitute a frame when translated and modulated according to this global lattice, the time step $a$ must be set with the hop size assigned to the smallest window frame. On the other hand, as the FFT of a discrete signal has the same number of points of the signal itself, the frequency step $b$ has to be the FFT size of the largest window analysis: for the smaller ones, a zero-padding is performed.\\

Each signal segment identifies a time-frequency rectangle $G$ for the entropy evaluation: the horizontal edge is the time interval of the considered segment, while the vertical one is the whole frequency lattice. For each spectrogram, the rectangle $G$ defines a subset of coefficients belonging to $G$ itself. The $|L|$ different subsets do not correspond to the same part of signal, as windows have different time supports. Therefore, a preliminary weighting of the signal has to be performed before the calculations of the local spectrograms: this step is necessary to balance the influence on the entropy calculus between coefficients which regard parts of signal shared or not shared by the different analysis frames.\\
After the pre-weighting, we calculate the entropy of every spectrogram as in \eqref{ren_ent_disc}. Having the $|L|$ entropy values corresponding to the different local spectrograms, the sparsest local analysis is defined as the one with minimum Rényi entropy: the window associated to the sparsest local analysis is chosen as best window at all the time points contained in $G$.\\
The global time adapted analysis of the signal is finally realized by opportunely assembling the slices of local sparsest analyses: they are obtained with a further spectrogram calculation of the unweighted signal, employing the best windows selected at each time point.\\

\begin{figure}
\centerline{\includegraphics[width=13.5cm]{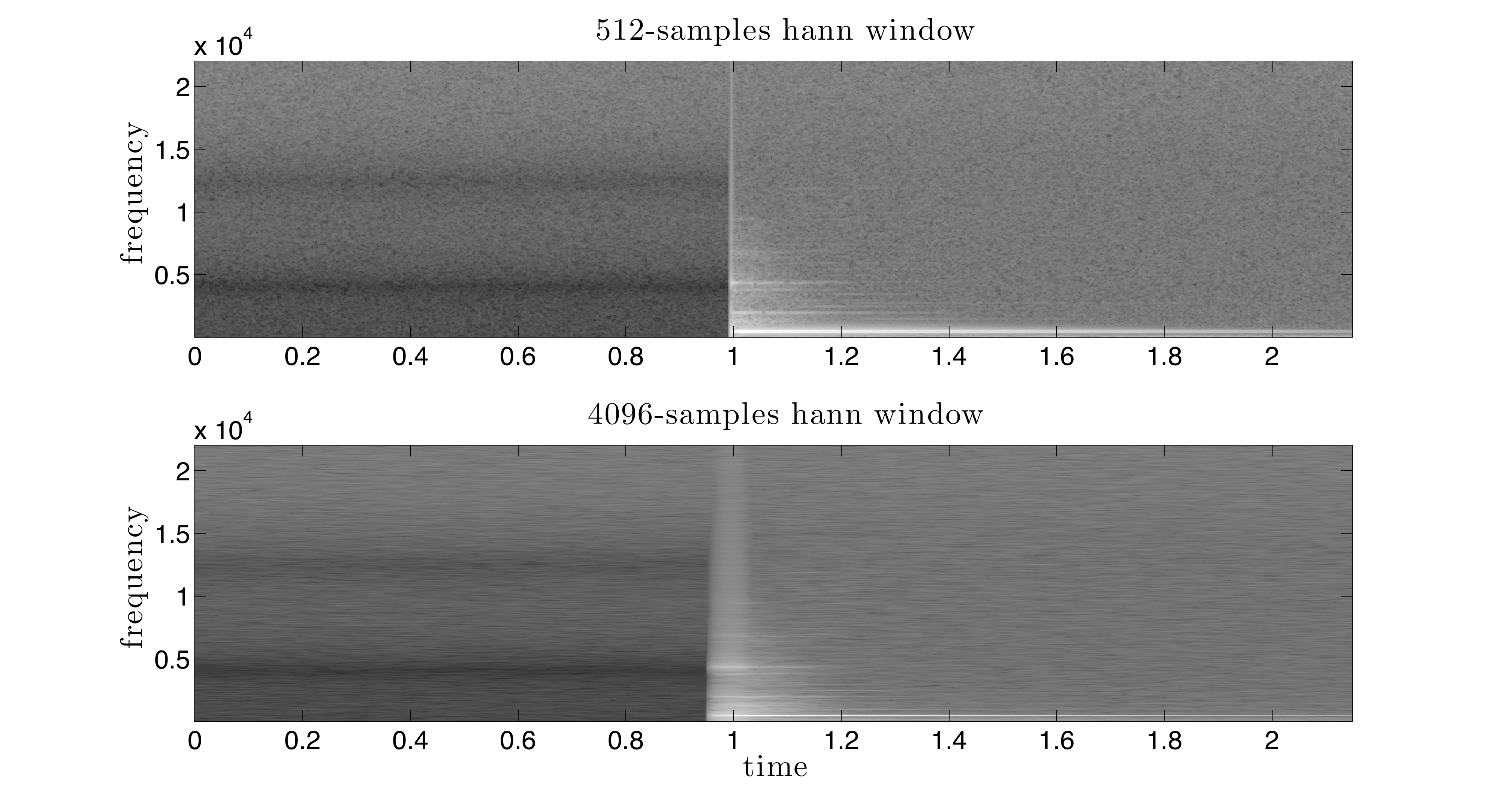}}
%\includegraphics[width=2in,bb = 3 257 607 534]{figures/marimba_B4_a07_o075_gray_1_2.pdf}
% The bounding box is set manually in this example. Useful for some .pdf figures.
\caption{\label{fig:marimba_1_2}{\it Two different spectrograms of a B4 note played by a marimba, with Hanning windows of sizes 512 (top) and 4096 (bottom) samples.}}
\end{figure}

\begin{figure}
\centerline{\includegraphics[width=13.5cm]{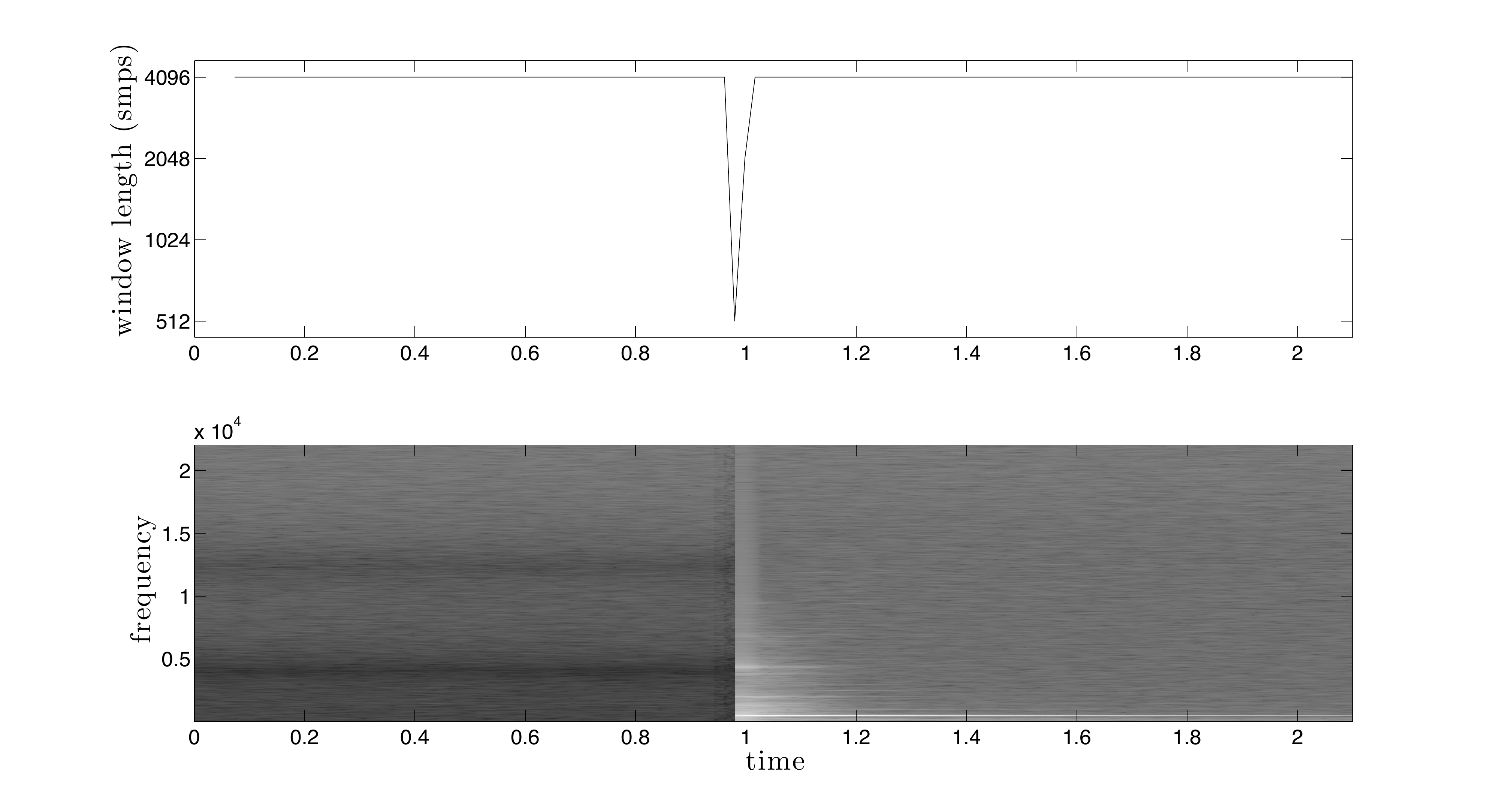}}
\caption[Figure 1]{\label{fig:marimba_B4_v1} \it Example  of an adaptive analysis performed by the first version of our algorithm with four Hanning windows of different sizes (512, 1024, 2048 and 4096 samples) on a B4 note played by a marimba: on top, the best window chosen as a function of time; at the bottom, the adaptive spectrogram. The entropy order is $\alpha = 0.7$ and each analysis segment contains twenty-four analyses frames with a sixteen-frames overlap between consecutive segments.}
\end{figure}

In figure \ref{fig:marimba_B4_v1} we give an example of an adaptive analysis performed by our first algorithm with four Hanning windows of different sizes on a real instrumental sound, a B4 note played by a marimba: this sound combines the need for a good time resolution at the strike with that of a good frequency resolution on the harmonic resonance. This is fully provided by the algorithm, as shown in the adaptive spectrogram at the bottom of the figure \ref{fig:marimba_B4_v1}.  Moreover, we see that the pre-echo of the analysis at the bottom of figure \ref{fig:marimba_1_2} is completely removed in the adapted spectrogram.\\ 

\begin{figure}
\centerline{\includegraphics[width=11.5cm]{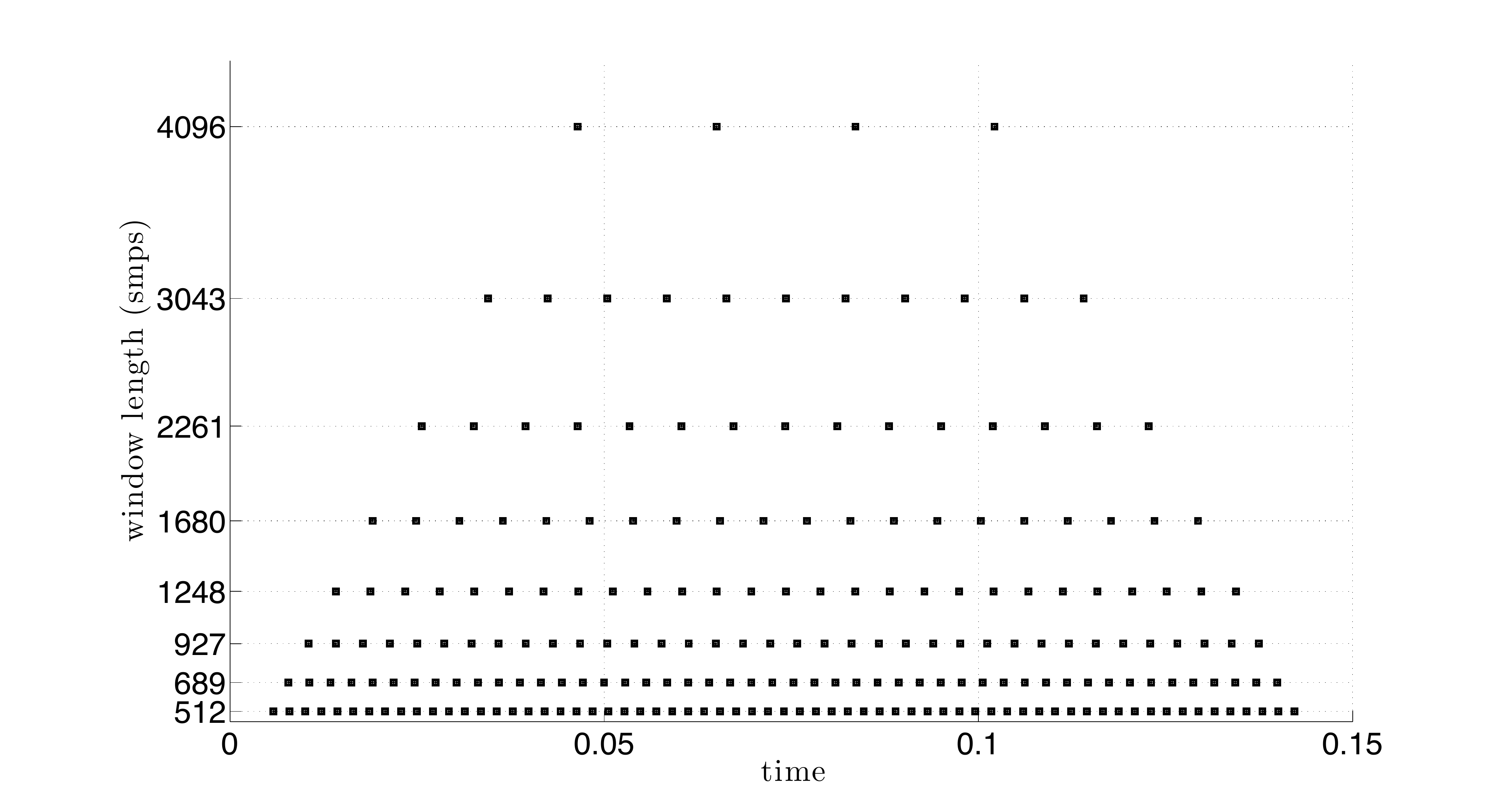}}
\caption{\label{fig:centers_var} {\it An analysis segment: time locations of the Heisenberg boxes associated to the multi-frame used in the second version of our algorithm.}}
\end{figure}

The main difference in the second version of our algorithm concerns the individual frames composing the multi-frame, which have the same frequency step $b$ but different time steps $\{a_l: l\in L\}$: the smallest and largest window sizes are given as parameters together with $|L|$, the number of different windows composing the multi-frame, and the global overlap needed for the analyses. The algorithm fixes the intermediate sizes so that, for each signal segment, the different frames have the same overlap between consecutive windows, and so the same redundancy.\\ 
This choice highly reduces the computational cost by avoiding unnecessary small hop sizes for the larger windows, and as we have observed in the previous section it does not affect the entropy evaluation. Such a structure generates an irregular time disposition of the multi-frame elements in each signal segment, as illustrated in figure \ref{fig:centers_var}; in this way we also avoid the problem of unshared parts of signal between the systems, but we still have a different influence of the boundary parts depending on the analysis frame: the beginning and the end of the signal segment have a higher energy when windowed in the smaller frames. This is avoided with a preliminary weighting: the beginning and the end of each signal segment are windowed respectively with the first and second half of the largest analysis window.\\
 As for the first implementation, the weighting does not concern the decomposition for re-synthesis purpose, but only the analyses used for entropy evaluations. 
%\begin{figure}
%\centering
%\includegraphics[height=6.2cm]{figures/frame_centers_var.pdf}
%\caption[Figure 1]{Time-frequency centers of the Heisenberg boxes associated to the multi-frame used in the second of our algorithms.}\label{fig4}
%\end{figure}
After the pre-weighting, the algorithm follows the same steps described above: calculation of the $|L|$ local spectrograms, evaluation of their entropy, selection of the window providing minimum entropy, computation of the adapted spectrogram with the best window at each time point, thus creating an analysis with time-varying resolution and hop size.\\

\begin{figure}
\centerline{\includegraphics[width=13.5cm]{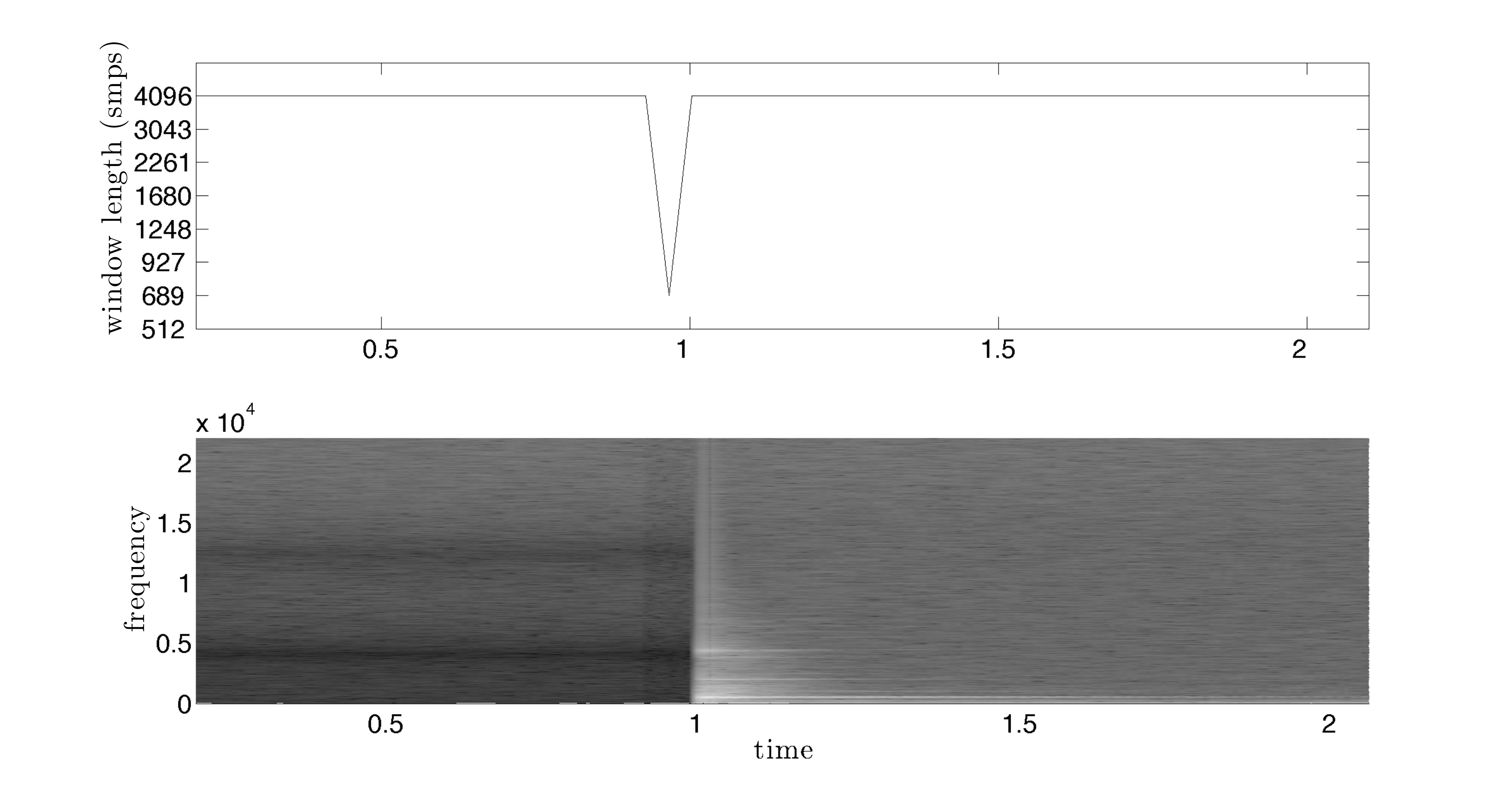}}
%\includegraphics[width=2in,bb = 3 257 607 534]{figures/marimba_B4_a07_o075_gray_1_2.pdf}
% The bounding box is set manually in this example. Useful for some .pdf figures.
\caption{\label{fig:marimba_B4_v2}{\it Example of an adaptive analysis performed by the second version of our algorithm with eight Hanning windows of different sizes from 512 to 4096 samples, on a B4 note played by a marimba sampled at 44.1kHz: on top, the best window chosen as a function of time; at the bottom, the adaptive spectrogram. The entropy order is $\alpha = 0.7$ and each analysis segment contains four frames of the largest window analysis with a two-frames overlap between consecutive segments.}}
\end{figure}

In figure \ref{fig:marimba_B4_v2} we give a first example of an adaptive analysis performed by the second version of our algorithm with eight Hanning windows of different sizes: the sound is still the B4 note of a marimba, and we can see that the two versions give very similar results. Thus, if the considered application does not specifically ask for a fixed hop size of the overall analysis, the second version is preferable as it highly reduces the computational cost without affecting the best window choice.\\

\begin{figure}
\centerline{\includegraphics[width=13.5cm]{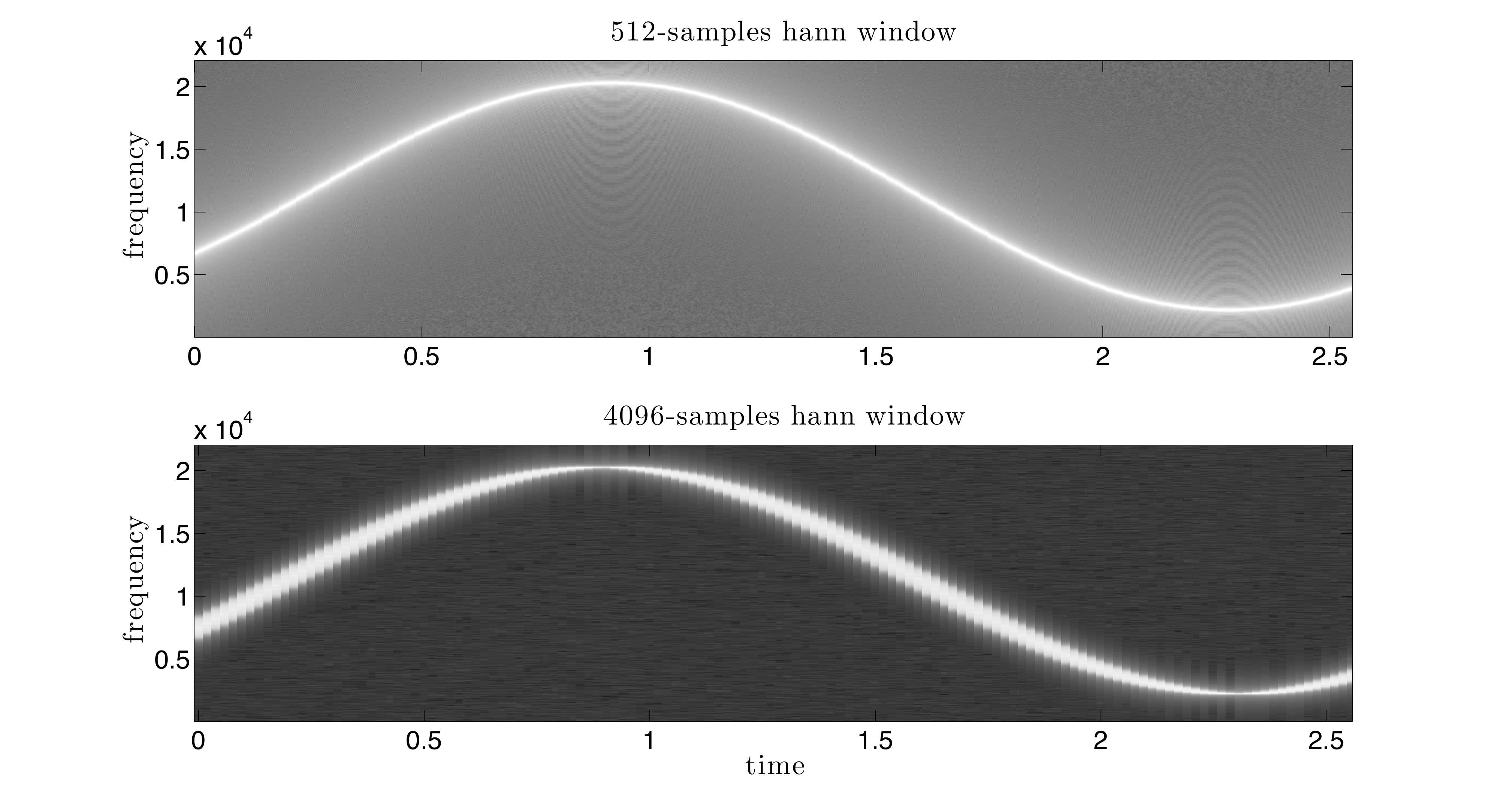}}
\caption{\label{fig:sin_sin_specs}{\it Two different spectrograms of a sinusoid with sinusoidal frequency modulation, with Hanning windows of sizes 512 (top) and 4096 (bottom) samples. }}
\end{figure}

\begin{figure}
\centerline{\includegraphics[width=13.5cm]{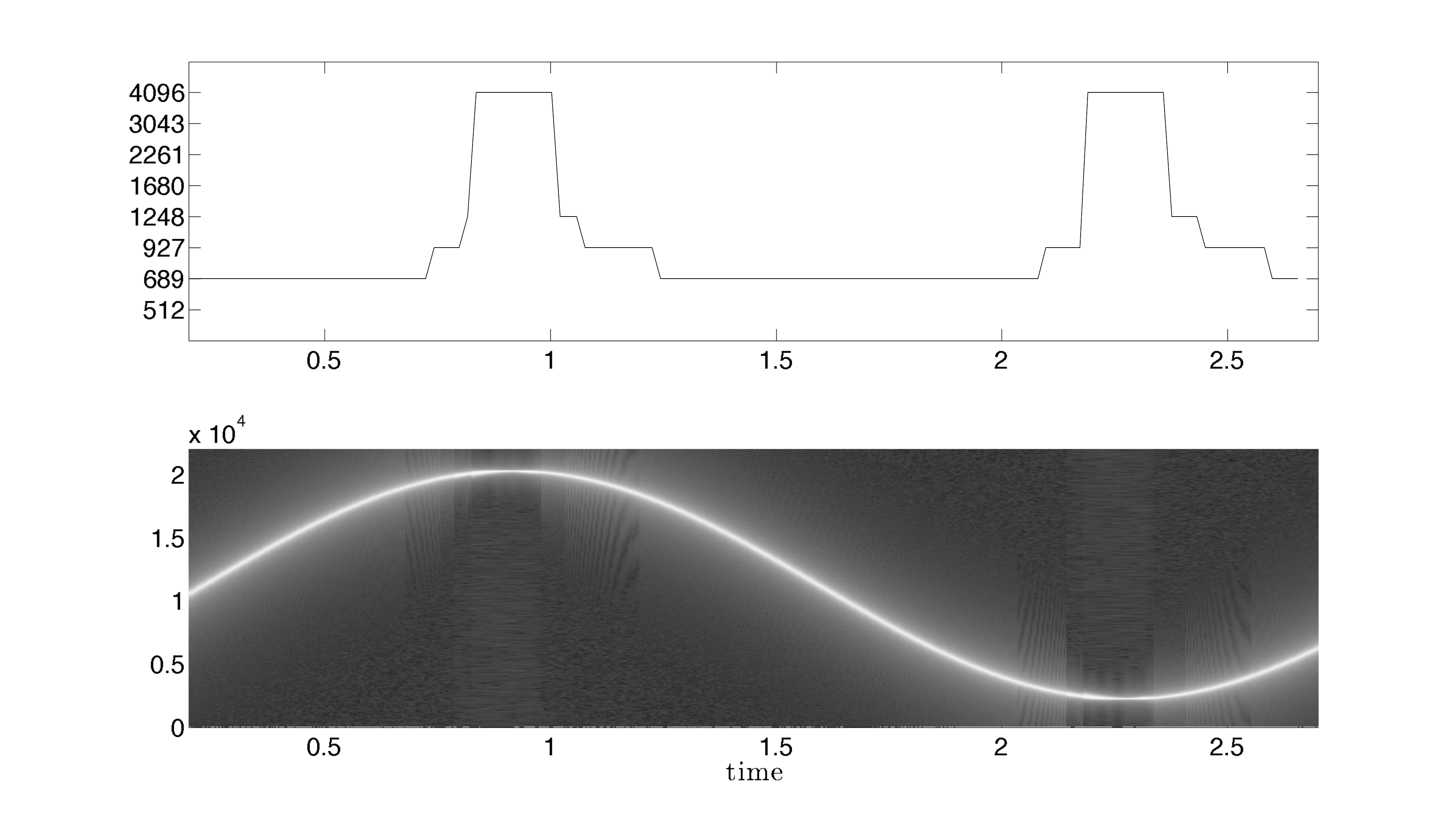}}
\caption{\label{fig:sin_sin}{\it Example of an adaptive analysis performed by the second version of our algorithm with eight Hanning windows of different sizes from 512 to 4096 samples, on a sinusoid with sinusoidal frequency modulation synthesized at 44.1 kHz: on top, the best window chosen as a function of time; at the bottom, the adaptive spectrogram. The entropy order is $\alpha = 0.7$ and each analysis segment contains four frames of the largest window analysis with a three-frames overlap between consecutive segments.}}
\end{figure}

In figure \ref{fig:sin_sin} we give a second example with a synthetic sound, a sinusoid with sinusoidal frequency modulation: as figure \ref{fig:sin_sin_specs} shows, a small window is best adapted where the frequency variation is fast compared to the window length; on the other hand,  the largest window is better where the signal is almost stationary.

%Different degree of continuity in the adapted spectrogram can be required by the user specifying the duration and overlap of the analysis segment: for not perfectly stationary signals, enlarging the duration increases the uniformity between the observed densities along the consecutive segments, thus smoothing within each analysis frame the entropy evaluation along time. As our algorithm defines best local resolutions according to a relative threshold on the previous minimal entropy value, such a smoothing decreases the probability of changes among different resolutions in the global adapted analysis.\\

\subsection{Re-synthesis Method}

The re-synthesis method introduced in \cite{GL84} gives a perfect reconstruction of the signal as a weighted expansion of the coefficients of its STFT in the original analysis frame. Let $S_f[n,k]$ be the STFT of a signal $f$ with window function $h$ and time step $a$; fixing $n$, through an iFFT we have a windowed segment of $f$
\beq\label{windowed_sig} f_h(n,l) = h(na-l)f(l)~,\eeq
whose time location depends on $n$. An immediate perfect reconstruction of $f$ is given by
\beq\label{perf_rec} f(l) = \frac{\sum_{n = -\infty}^{+\infty} h(na - l)f_h(n,l)}{\sum_{n = -\infty}^{+\infty}h^2(na - l)}~.\eeq
In our case, after the automatic selection step we dispose of a temporal sequence with the best windows at each time position; in the first version we have a fixed hop for all the windows, in the second one every window has its own time step. In both the cases we have thus reduced the initial multi-frame to a nonstationary Gabor frame: we extend the same technique of \eqref{perf_rec} using a variable window $h$ and time step $a$ according to the composition of the reduced multi-frame, obtaining a perfect reconstruction as well. The interest of \eqref{perf_rec} is that the given distribution does not need to be the STFT of a signal: for example, a transformation $S^*[n,k]$ of the STFT of a signal could be considered. In this case, \eqref{perf_rec} gives the signal whose STFT has minimal least squares error with $S^*[n,k]$.\\
As seen by the equations \eqref{nonstat_frame_def} and \eqref{dual_frame_nonst}, the theoretical existence and the mathematical definition of the canonical dual frame for a nonstationary Gabor frame like the one we employ has been provided \cite{JBD09}: it is thus possible to define the whole analysis and re-synthesis framework within the Gabor theory. We are at present working on the interesting analogies between the two approaches, to establish a unified interpretation and develop further extensions.
%
%\begin{figure}
%\centering
%\includegraphics[height=6.2cm]{figures/marimba_B4_a07_o075_gray_3_4.pdf}
%\caption[Figure 1]{Example  of an adaptive analysis performed by our algorithm with four Hanning windows of different sizes (512, 1024, 2048 and 4096 samples) on a B4 note played by a marimba: on top, the best window chosen as a function of time; at the bottom, the adaptive spectrogram with $\alpha = 0.7$.}\label{fig3}
%\end{figure}
\section{Conclusions}
We have presented an algorithm for time-adaptation of the spectrogram resolution, which can be easily integrated in existent framework for analysis, transformation and re-synthesis of an audio signal: the adaptation is locally obtained through an entropy minimization within a finite set of resolutions,  which can be defined by the user or left as default. The user can also specify the time duration and overlap of the analysis segments where entropy minimization is performed, to privilege more or less discontinuous adapted analyses.\\
Future improvements of this method will concern the spectrogram adaptation in both time and frequency dimensions: this will provide a decomposition of the signal in several layers of analysis frames, thus requiring an extension of the proposed technique for re-synthesis.

\end{document}